\documentclass[conference]{IEEEtran}
\IEEEoverridecommandlockouts
\usepackage{cite}
\usepackage{amsmath,amssymb,amsfonts}
\usepackage{algorithmic}
\usepackage{graphicx}
\usepackage{textcomp}
\usepackage{xcolor}
\usepackage{authblk}

\usepackage{booktabs}
\usepackage{gensymb}
\usepackage{subcaption}
\usepackage{balance}
\usepackage{hyperref}

\def\BibTeX{{\rm B\kern-.05em{\sc i\kern-.025em b}\kern-.08em
    T\kern-.1667em\lower.7ex\hbox{E}\kern-.125emX}}
\begin{document}

\newcommand{\eg}{\emph{e.g.}~}
\newcommand{\etal}{\emph{et al.}~}
\newcommand{\ie}{\emph{i.e.}~}
\newcommand{\wrt}{\emph{w.r.t.}~}

\newcommand{\qj}[1]{\textcolor{blue}{[\textbf{QJ}: #1]}}
\newcommand{\xr}[1]{\textcolor{purple}{[\textbf{XR}: #1]}}
\newcommand{\modify}[1]{\textcolor{red}{#1}}
\renewcommand{\thefootnote}{}
\title{Supervised Domain Adaptation for Recognizing Retinal Diseases from Wide-Field Fundus Images 
\thanks{This research was supported by the National High Level Hospital Clinical Research Funding (2022-PUMCH-C-61), NSFC (62172420),  Fund for Building World-Class Universities (Disciplines) of Renmin University of China, and Public Computing Cloud, Renmin University of China.}
}

\author[1,2]{Qijie Wei}
\author[3]{Jingyuan Yang}
\author[2]{Bo Wang}
\author[2]{Jinrui Wang}
\author[2]{Jianchun Zhao}
\author[3]{Xinyu Zhao}
\author[4]{Sheng Yang}

\author[4]{\\Niranchana Manivannan}
\author[3]{Youxin Chen}
\author[2]{Dayong Ding}
\author[1]{Jing Zhou}
\author[1]{Xirong Li$^{*}$}

\affil[1]{Renmin University of China, Beijing, China}
\affil[2]{Vistel, Beijing, China}
\affil[3]{Peking Union Medical College Hospital, Beijing, China}
\affil[4]{Carl Zeiss Meditec, Dublin, United States}

\affil[ ]{\href{https://github.com/ruc-aimc-lab/CdCL}{https://github.com/ruc-aimc-lab/CdCL}}
\maketitle

\begin{abstract}
   This paper addresses the emerging task of recognizing multiple retinal diseases from wide-field (WF) and ultra-wide-field (UWF) fundus images. For an effective use of existing large amount of labeled color fundus photo (CFP) data and the relatively small amount of WF and UWF data, we propose a supervised domain adaptation method named Cross-domain Collaborative Learning (CdCL). Inspired by the success of fixed-ratio based mixup in unsupervised domain adaptation, we re-purpose this strategy for the current task. Due to the intrinsic disparity between the field-of-view of CFP and WF/UWF images, a scale bias naturally exists in a mixup sample that the anatomic structure from a CFP image will be considerably larger than its WF/UWF counterpart. The CdCL method resolves the issue by Scale-bias Correction, which employs Transformers for producing scale-invariant features. As demonstrated by extensive experiments on multiple datasets covering both WF and UWF images, the proposed method compares favorably against a number of competitive baselines.

\end{abstract}


\footnotetext{$^{*}$Corresponding author: Xirong Li (xirong@ruc.edu.cn)}
\section{Introduction}
\label{intro}

\begin{figure}[tb!]
    \centerline{
        \subcaptionbox{CFP, 45\degree}{
            \label{fig:cfp}
            \includegraphics[height=0.3\columnwidth]{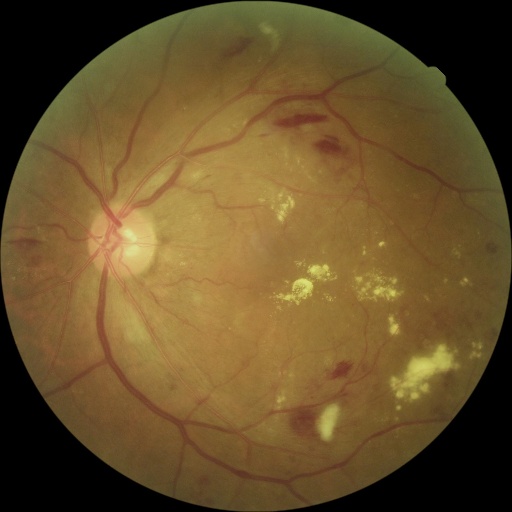}}
        \subcaptionbox{WF, 133\degree}{
            \label{fig:wf}
            \includegraphics[height=0.3\columnwidth]{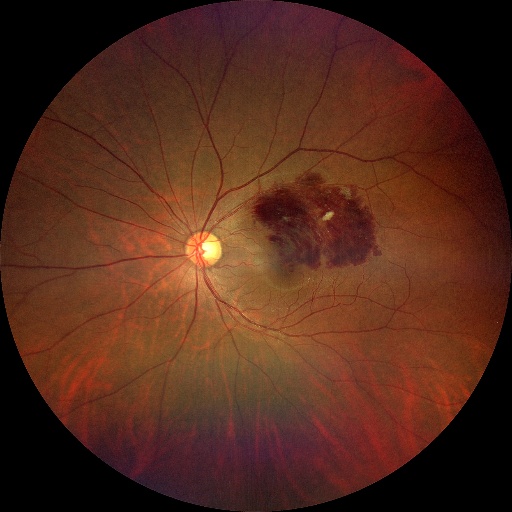}}
        \subcaptionbox{UWF, 200\degree}{
            \label{fig:uwf}
            \includegraphics[height=0.3\columnwidth]{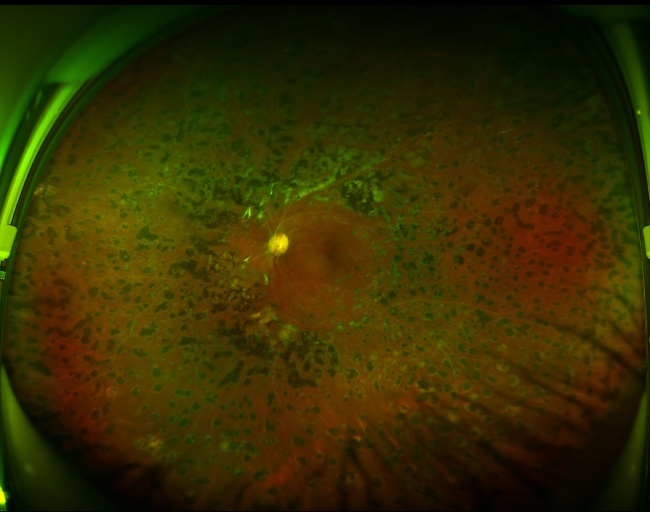}}}
    \vspace{1mm}    
    \centerline{
        \subcaptionbox{0.3$\times$(a)+0.7$\times$(b)}{
            \label{fig:wf_mix}
            \includegraphics[height=0.44\columnwidth]{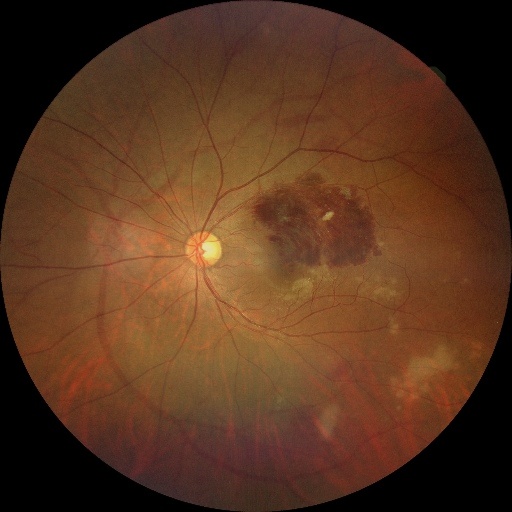}}
        
        \subcaptionbox{0.3$\times$(a)+0.7$\times$(c)}{
            \label{fig:uwf_mix}
            \includegraphics[height=0.44\columnwidth]{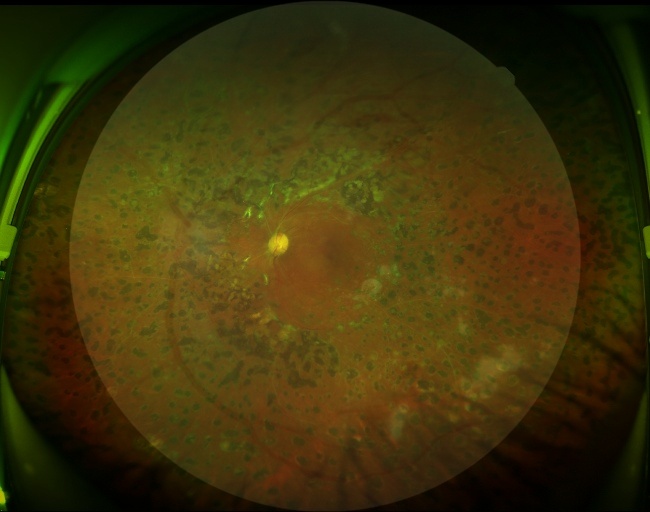}}}
            
    \caption{\textbf{Fundus images with varied FoV and their mixups}. (a) CFP  of 45 degree FoV taken by Canon CR-2. (b) WF image of 133 degree FoV taken by Zeiss Clarus500. (c) UWF image of 200 degree FoV taken by Optos Daytona which only has red and green channel. (d) Mixup of WF and CFP images. (e) Mixup of UWF and CFP images. Best viewed digitally.
    }
    \label{fig:example}
    \vspace{-0.5cm}
\end{figure}

Wide-field (WF) and ultra-wide-field (UWF) fundus imaging are playing an increasingly important role in fundus condition assessment and early diagnosis of retinal diseases \cite{patel2020ultra}. Compared to traditional color fundus photography (CFP), WF/UWF images have substantially larger field of view (FoV), see Fig. \ref{fig:example}, making it possible for visualization of pathological alterations in peripheral retina \cite{nagiel2016ultra}. Not surprisingly, deep learning methods for retinal disease recognition from such larger-FoV images are being actively developed \cite{abitbol2022deep,sun2022deep,zhang2021development,oh2021early}.

\begin{figure}[htb!]
    \includegraphics[height=0.35\textwidth]{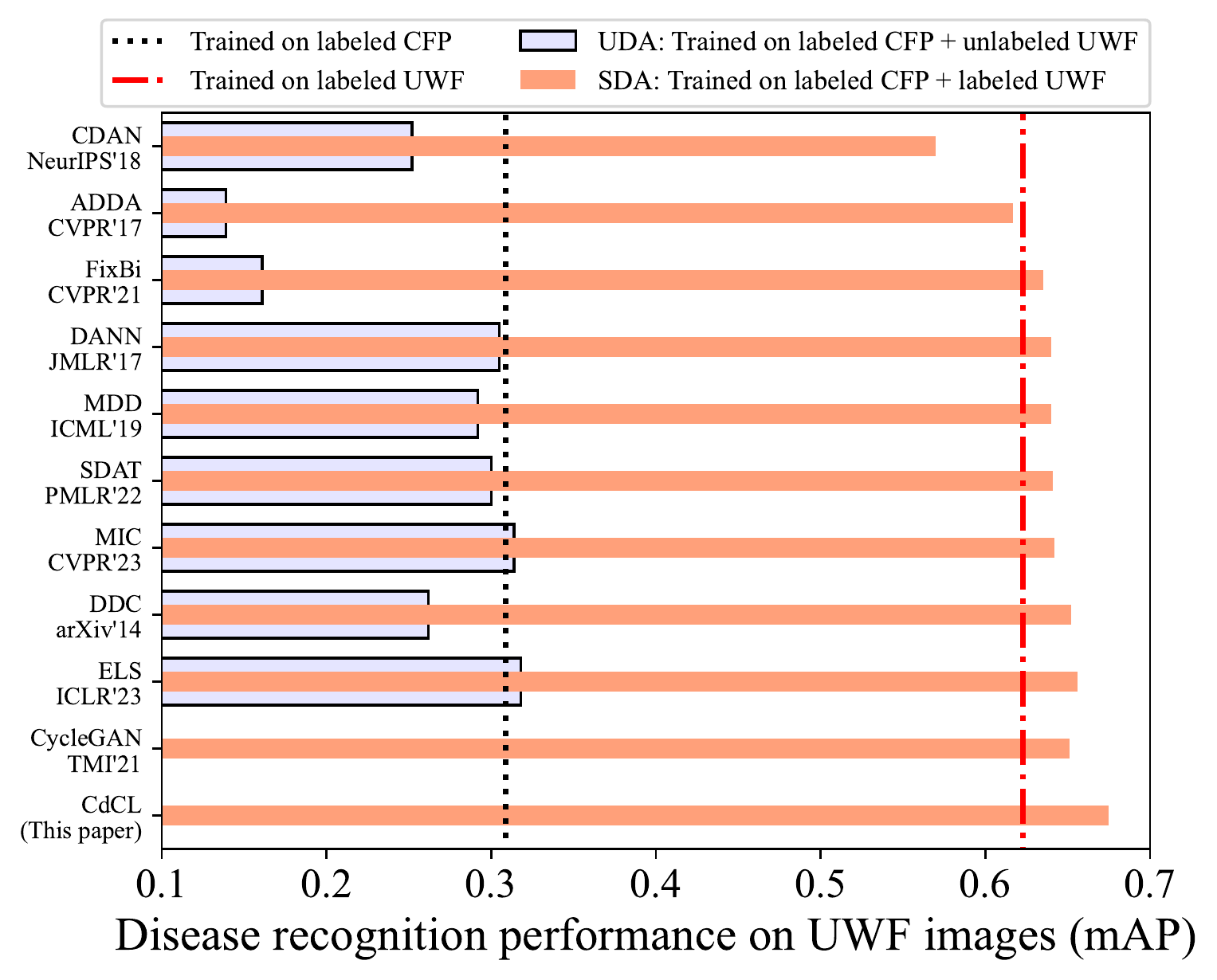}
    \caption{\textbf{Performances of different methods on UWF image classification}. CFP is source domain and UWF is target domain. The black dotted line and red dash-dotted line refer to the performances of models trained exclusively on CFP data or UWF data, respectively.}
    \label{fig:uda}
    \vspace{-0.5cm}
\end{figure}

In contrast to the CFP domain, wherein quite a few labeled datasets are publicly available \cite{RFMiD,kaggle,lesion-net,FGADR,IDRID} thanks to its long-lasting research, annotated WF/UWF images are in short supply. 
Recently, there have been few novel attempts to exploit the rich CFP data for UWF image classification \cite{ju2021leveraging,bai2022unsupervised}. Ju \etal propose to train several Cycle GANs \cite{cyclegan} to generate UWF images from a given CFP image. Labels associated with the given CFP image is then used as pseudo labels for the generated UWF images. Then, a convolutional neural network (CNN) pre-trained on a limited set of labeled UWF images is used to select high-quality samples from the generated data \cite{ju2021leveraging}. For balancing the reliability and the informativeness of the samples, such a selection is nontrivial. 
Bai \etal propose an unsupervised lesion-aware transfer learning method for diabetic retinopathy (DR) grading in UWF images \cite{bai2022unsupervised}. Specifically, domain-invariant adversarial learning is applied to both the underlying CFP-based lesion segmentation model and DR grading model to improve the cross-domain usability among them. Nonetheless, adversarial learning can be tricky by itself and the proposed method requires additional lesion segmentation annotations.

In fact, although several state-of-the-art unsupervised domain adaptation (UDA) methods that utilize adversarial learning or feature discrepancy to align features between source domain and target domain have achieved in certain tasks \cite{DDC,DANN,ADDA,MDD,SDAT,ELS}, they fail in bridging the domain gap between CFP and UWF images. As shown in Fig.~\ref{fig:uda}, compared to model trained exclusively on CFP data, the performance of the UDA methods do not benefit from their feature alignment. This highlights the challenge of directly accomplishing feature alignment between CFP and UWF images due to the distinct domain gap, primarily in both FoV and color.

On the other hand, some domain adaptation methods, as manifested by FixBi \cite{FixBi}, uses no adversarial learning. For the cross-domain use of labeled images from a source domain, FixBi uses a fixed-ratio based mixup strategy, where a weighted combination of a source-domain image and a target-domain image is used as a new training image. In particular, the source-domain image is assigned with a small weight, say 0.3, so the new image will be visually more close to the target domain. Such a simple strategy is found to be effective for cross-domain image classification \cite{FixBi}. However, we argue that directly applying this strategy in the current context is problematic. As we have noted, the FoV of a CFP image is noticeably smaller than that of our target domain, \ie WF or UWF images. Given a CFP image and a UWF image of close size, the anatomic structure from the CFP image will be considerably larger than that from the UWF when mixing the two images up, see Fig. \ref{fig:example}. Consequently, a scale bias will be systemically introduced when learning from such mixups. One might consider remedying the issue, either by downsizing the CFP image or by upsizing the UWF image. The former will inevitably cause information loss, while the latter is also questionable. Bounded by GPU memory, using a larger input means reducing the batch size that in turn affects adversely network optimization. 

While the field of domain adaptation mainly focuses on unsupervised domain adaptation (UDA) problems, where the target domain lacks annotation, we argue that in our case, a relatively small amount of labeled training examples can be available for the target domain. As shown in Fig.~\ref{fig:uda}, the performance of existing methods significantly improves when labels are available in the target domain. Hence, we concentrate on supervised domain adaptation (SDA), a more practical approach in our problem, where both the source domain and the target domain have annotations.

\begin{figure*}[htb!]
    \centerline{\includegraphics[width=1.\textwidth]{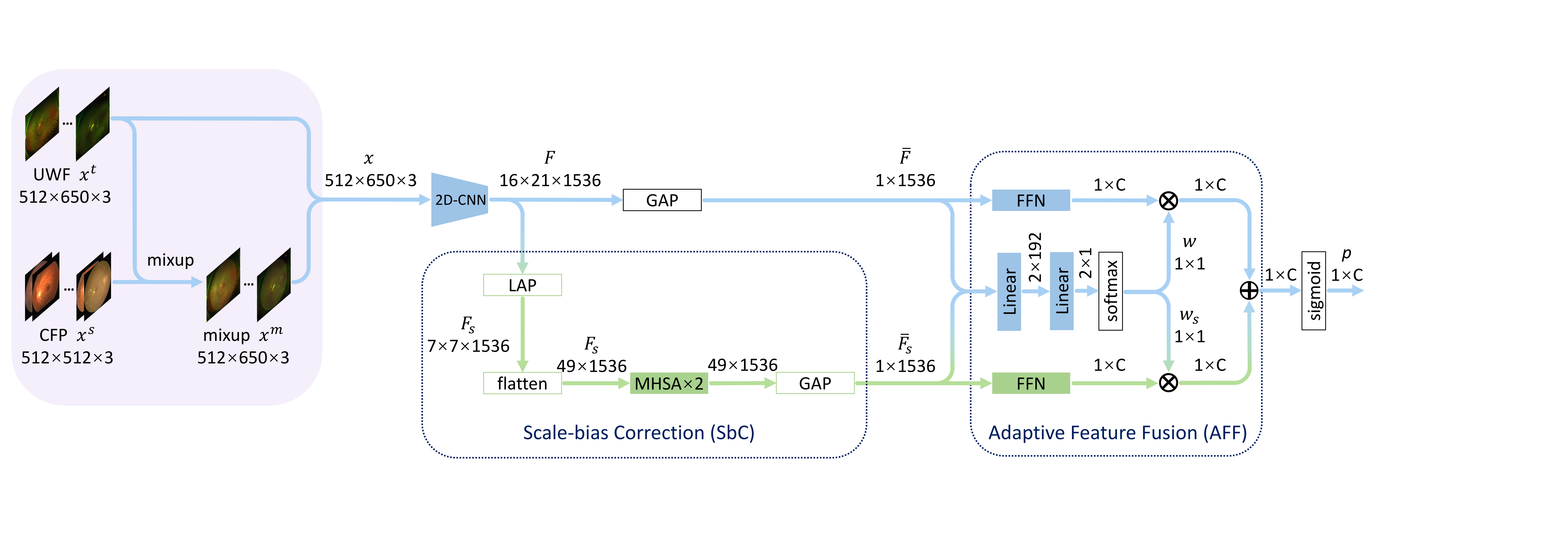}}
    \caption{\textbf{Proposed \texttt{CdCL} method for training a UWF image based retinal disease recognition network}. 
    }
    \label{fig:model}
    \vspace{-0.5cm}
\end{figure*}

Towards a more effective use of the rich CFP data for WF/UWF image based recognition of multiple retinal diseases, we propose in this paper Cross-domain Collaborative Learning (\texttt{CdCL}), see Fig. \ref{fig:model}. We follow the good practice of FixBi, using CFP data with mixup. To attack the scale-bias issue, we propose Scale-bias Correction (SbC), which can be effectively implemented with Transformers \cite{transformers}. Furthermore, an Adaptive Feature Fusion (AFF) module is adopted for multi-scale feature fusion.

Our main contributions are:

$\bullet$ We propose \texttt{CdCL}, a novel framework for cross-domain collaborative learning between CFP and WF/UWF images. The proposed framework resolve the scale-bias between CFP and WF/UWF images through Scale-bias Correction module.

$\bullet$ We make the first attempt to develop a unified framework that utilizes CFP data for classification in various fundus image types, namely WF and UWF images. Comprehensive experiments on multiple datasets covering both WF and UWF images justify the viability of the proposed method.

\section{Related Work} \label{rel}

\textbf{Cross-domain learning for UWF images}. 
Attempts have been made to build deep learning models for diseases classification in UWF images with the assistance of large scale CFP dataset. 
Ju \etal \cite{ju2021leveraging} proposed a generative adversarial network (GAN) \cite{GAN} based framework to leverage CFP data for UWF image classification. Several cycle GANs \cite{cyclegan} were built to generate UWF images from CFP. The generated UWF images were selected according to their quality which was measured by a pretrained preliminary UWF classification model. Subsequently, labels from CFP images were assigned to the corresponding selected generated UWF images and the final UWF classification model was then trained on these UWF images. 
Bai \etal \cite{bai2022unsupervised} proposed a lesion-aware network for Diabetic Retinopathy (DR) grading in UWF images. Domain adversarial learning was applied between CFP and UWF images for both lesion segmentation model and DR grading model.
In contrast to \cite{ju2021leveraging} that relies on additional trained Cycle GANs and \cite{bai2022unsupervised} which requires extra lesion segmentation annotations on CFP, our proposed framework utilizes CFP data directly and requires no further annotations on CFP data.
 
\textbf{Unsupervised domain adaptation}. 
Unsupervised domain adaptation (UDA) aims at adapting a model trained on a labeled source domain to an unlabeled target domain. Current UDA methods rely either on domain-invariant feature learning  \cite{DDC,ADDA,DANN,MDD,CDAN,SDAT,ELS} or on self-supervised learning \cite{FixBi,MIC,tomm2021-ude}.


Domain-invariant feature learning is typically achieved through feature discrepancy minimization or adversarial training.
Tzeng \etal \cite{DDC} and Zhang \etal \cite{MDD} proposed different measurements of feature discrepancy between source domain and target domain. The feature discrepancy was minimized during training.
Tzeng \etal \cite{ADDA} and Ganin \etal \cite{DANN} employed adversarial training, where a domain discriminator was built to differentiate features from the source and target domains while its corresponding feature extractor aimed to produce domain-invariant features to deceive the domain discriminator.
Long \etal \cite{CDAN} proposed a conditional domain discriminator that took both feature representations and classifier predictions as inputs simultaneously. 
Based on previous adversarial training methods, Rangwani \etal \cite{SDAT} proposed a Smooth Task Loss to increase the smoothness during adversarial training, resulting in better generalization in the target domain. 
Zhang \etal \cite{ELS} demonstrated that changing the domain label into soft label could lead to better performance in the target domain.

Self-supervised learning approaches  rely on pseudo labels.
Na \etal \cite{FixBi} proposed a two-stream architecture with a mixup strategy. Predictions of each stream are used as pseudo labels for the other stream.
Hoyer \etal \cite{MIC} proposed Masked Image Consistency, where the predictions of the masked images were treated as pseudo labels. Wang \etal \cite{tomm2021-ude} resort to knowledge distillation to exploit pseudo labels generated by domain-specific models.


However, current UDA methods exhibit poor performance in UWF images, as depicted in Fig.~\ref{fig:uda}, primarily due to the distinct domain gap between CFP and WF/UWF images in both FoV and color. To address this issue, our \texttt{CdCL} utilizes the mixup strategy to reduce the disparity in color and resolves the scale-bias caused by the disparity in FoV.

\section{Proposed Method}
\label{met}

Let $x$ be a specific fundus image, associated with a binary label vector $y$ that indicates the relevance of the image \wrt a pre-defined set of $C$ fundus diseases. Suppose we have access to a large set of $n^s$ labeled images $D^s=\{(x_i^s, y_i^s), i=1, 2, ..., n^s\}$ from the source (CFP) domain and a relatively small set of $n^t$ labeled images $D^t=\{(x_i^t, y_i^t), i=1, 2, ..., n^t\}$ from the target (WF or UWF) domain, with $n^t \ll n^s$. By exploiting both $D^s$ and $D^t$, we aim for a better target-domain classifier. To that end, we propose Cross-domain Collaborative Learning (\texttt{CdCL}).

\subsection{Cross-domain Sample Construction} \label{ssec:mixup}

We adopt fixed-ratio based mixup \cite{FixBi} to construct cross-domain training samples. In particular,  by mixing $(x^s, y^s)$ and $(x^t, y^t)$ sampled randomly from $D^s$ and $D^t$, a new training sample $(x^m, y^m)$ is generated as 
\begin{equation} \label{eq:mixup}
\left\{ \begin{array}{l}
x^m = \lambda x^t + (1-\lambda) x^s, \\
y^m = \lambda y^t + (1-\lambda) y^s,
\end{array}
\right.
\end{equation}
where $\lambda \in (0,1)$ is a fixed ratio. To make the new sample visually close to the target domain, we use $\lambda$ of 0.7. Our mini-batch for network training will have $b$ target-domain samples and their mixup counterparts of the same amount, thus balanced. The mixup strategy allows the source-domain data to easily collaborate with the target-domain data. 




\subsection{Network for CdCL} \label{ssec:network} 

Given a $w \times h$ input image, let it be $x^t$ or $x^m$, our network has a regular 2D-CNN (EfficientNet-B3 \cite{efficientnet}) to extract a downsized $d$-channel feature map $F$ of size $\frac{w}{32}\times\frac{h}{32}\times d$. In a single-domain scenario,  global average pooling (GAP) is often used to convert $F$ to a $d$-dimensional feature vector $\bar{F}$. The vector then goes through a feedforward  network (FFN) to produce a $C$-dimensional probability vector $p$ \wrt the classes. In the current cross-domain scenario, however, the above pipeline is questionable as the feature map has been scale-biased. We thus introduce scale-bias correction (SbC). 

\textbf{Scale-bias correction}. For scale-invariant feature extraction, local feature interaction in a global context is crucial. For instance, although the optic disc from $x^s$ is noticeably larger than the optic disc from $x^t$, the feature map concerning $x^m$ shall have similar values describing the two optic discs. 
We therefore utilize Transformers \cite{transformers}, designed to exploit inter-feature relationships in a global context. 
To achieve this, we first adopt a local average pooling (LAP) instead of commonly used GAP to reduce $F$ into a smaller feature map $F_s$ of size $w_s \times h_s \times d$ which represents $(w_s \times h_s)$ $d$-dimension local features. The kernel size of LAP is set to $\frac{w}{32 \times r}\times\frac{h}{32 \times r}$ and stride is half of the kernel size. The ratio $r \geq 1$ is a hyper-parameter controls the scope of local feature where larger $r$ refers to smaller kernel size and smaller scope of local feature.
After flattening, a sequence of $w_s \times h_s $ features are fed to a standard Transformer consisting of two multi-head self attention (MHSA) blocks. With mean pooling on the output sequence of the Transformer, a scale-bias corrected feature vector $\bar{F}_s$ is obtained.

\textbf{Adaptive feature fusion}. We shall not discard $\bar{F}$ as it may still be complementary to $\bar{F}_s$. Moreover, the former has a shorter path back to the backbone than the latter, making gradient back propagation more efficient and thus of benefit to backbone training \cite{mvssnet-pami}. 
In order to jointly exploiting the two features for producing the probability vector $p$, we perform adaptive feature fusion module as follows. Each feature is fed to a separate FFN to obtain a $C$-dimensional score vector. Meanwhile, the two features are stacked and fed into a lightweight attention block, which consists of two linear layers followed by softmax, to produce two feature-specific weights $w$ and $w_s$. Accordingly, the weighted average of the two score vectors is obtained and converted to $p$ by a sigmoid function for multi-label classification.

To sum up, our network computes $p$ for the given image $x$ as follows: 
\begin{equation} \label{eq:network}
\left\{ \begin{array}{ll}
F & \leftarrow \mbox{CNN}(x), \\
\bar{F} & \leftarrow\mbox{GAP}(F),\\
F_s & \leftarrow\mbox{LAP}(F), \\
\bar{F}_s & \leftarrow \mbox{GAP(MHSA*2(~flatten(}F_s\mbox{)~))},\\
w, w_s & \leftarrow \mbox{softmax(Linear*2(}[\bar{F}; \bar{F}_s]\mbox{)~)},\\
p & \leftarrow \mbox{sigmoid(}w\cdot \mbox{FFN}(\bar{F}) + w_s \cdot \mbox{FFN}(\bar{F}_s)). \\
\end{array}
\right.
\end{equation}
The network is trained by minimizing the BCE loss.

\section{Evaluation}
\label{exp}

\subsection{Experimental Setup}

\textbf{Data}.
Five datasets are used, see Table \ref{tab:dataset}, among which CFP-MD, WF-MD and UWF-MD are our private collections, while RFMiD \cite{RFMiD} and TOP$^1$\footnotetext{$^1$\href{https://github.com/DateCazuki/Fundus\_Diagnosis}{https://github.com/DateCazuki/Fundus\_Diagnosis}} are public.
CFP-MD, collected in our earlier research, consists of 31.6k color fundus photos taken by different fundus cameras including Canon CR-2 / CR-DGI, Topcon NW400 / TRC-50DX, KOWA Nonmyd 7 and ZEISS VISUCAM 244. WF-MD has 6.9k WF fundus images taken by ZEISS Clurus500, while UWF-MD has 6.5k UWF fundus images taken by Optos Daytona. Both WF-MD and UWF-MD were acquired from the Department of Ophthalmology, PUMCH$^2$\footnotetext{$^2$The study has complied with the Declaration of Helsinki.} in 2022. CFP-MD, WF-MD and UWF-MD have the following eight fundus diseases in common, all labeled by retinal specialists: Diabetic Retinopathy (DR), Retinal Vein Occlusion (RVO), Laser Photocoagulation (LP), Degenerative Myopia (DM), Retinal Detachment (RD), Tigroid Fundus(TF), Age Related Macular Degeneration (AMD) and Macular Epiretinal Membrane (MEM). Note that trying all possible combinations of the datasets is computationally prohibitive. So we conduct experiments in the following two groups: 1) \emph{Private group}: CFP-MD as the source domain, WF-MD / UWF-MD as the target domain, and 2) \emph{Public group}: RFMiD as the source domain, TOP as the target domain.

\begin{table}[htb!]
    \caption{\textbf{Our experimental data}. 
    We  evaluate models on the three WF / UWF datasets, \ie UWF-MD, WF-MD and TOP, so the val. / test splits of the two CFP datasets, \ie CFP-MD and RFMiD, are ommited. }
    \label{tab:dataset}
    \renewcommand{\arraystretch}{1.1}
    \centerline{
        \scalebox{0.68}{
            \begin{tabular}{@{}l rrr  rrrrrrrr@{}}
               \toprule
                \textbf{Dataset} & \textbf{Train}& \textbf{Val.} & \textbf{Test} & \textit{DR} & \textit{RVO} & \textit{LP} & \textit{DM} & \textit{RD} & \textit{TF} & \textit{AMD} & \textit{MEM} \\
                \midrule
                \emph{Private:} \\
                CFP-MD      &  31,623 & - & -   & 2,094       & 2,268        & 525         & 8,068       & 329         &  7,369      & 3,637        & 968          \\
                WF-MD       & 4,211 & 1,304 & 1,417 & 697         & 399          & 1,450       & 366         & 265         &  1,192      & 592          & 442          \\
                UWF-MD      & 3,962 & 1,299 & 1,304   & 620         & 164          & 1,307       & 572         & 448         &  806        & 133          & 95           \\ [2pt]
                \emph{Public:} \\
                RFMiD \cite{RFMiD}    & 1,920    & -   & -   & 595         & 157          & 79          & 149         & -           &  -          & 162          & 26           \\
                TOP     & 788 & 2,542 & 2,620 & 1,514      & 364     & -    & -           & 454        &  -          & 205          & -      \\
              \bottomrule
            \end{tabular}
        }
    }
    \vspace{-0.3cm}
\end{table}

\begin{table}[htb!]
    \caption{\textbf{Evaluation on UWF-MD}.}
    \label{tab:uwf-md}
    \renewcommand{\arraystretch}{1.1}
    \centerline{
        \scalebox{0.65}{
            \begin{tabular}{@{}lrrrrrrrrr@{}}
        \toprule
                \textbf{Method} & \textbf{mAP} & \textbf{DR} &  \textbf{RVO} & \textbf{LP} & \textbf{DM} & \textbf{RD} & \textbf{TF} & \textbf{AMD} & \textbf{MEM}\\
                \midrule
                \multicolumn{2}{@{}l}{\textit{Supervised baselines:}}\\
                \texttt{CFP} & 0.309$\pm$0.025 & 0.581 & 0.240 & 0.472 & 0.463 & 0.415 & 0.183 & 0.064 & 0.052 \\
                \texttt{UWF} & 0.622$\pm$0.018 & 0.726 & 0.498 & 0.927 & 0.868 & 0.840 & 0.620 & 0.340 & 0.161\\
                \texttt{UWF+} & 0.623$\pm$0.005 & 0.721 & 0.498 & 0.923 &0.883 & 0.843 & 0.611 & 0.309 & 0.198\\
                
                \midrule
                \multicolumn{2}{@{}l}{\textit{UDA baselines:}}\\
                \texttt{ADDA} \cite{ADDA}  & 0.140$\pm$0.024 & 0.202 & 0.027 & 0.254 & 0.256 & 0.098 & 0.244 & 0.021 & 0.013\\
                \texttt{FixBi} \cite{FixBi} & 0.161$\pm$0.023 & 0.065 & 0.050 & 0.544 & 0.215 & 0.090 & 0.288 & 0.021 & 0.015\\
                \texttt{CDAN} \cite{CDAN} & 0.252$\pm$0.024 & 0.411 & 0.151 & 0.432 & 0.657 & 0.066 & 0.255 & 0.032 & 0.015\\
                \texttt{DDC} \cite{DDC} & 0.262$\pm$0.006 & 0.468 & 0.329 & 0.509 & 0.481 & 0.097 & 0.157 & 0.042 & 0.014\\ 
                \texttt{MDD} \cite{MDD} & 0.292$\pm$0.007 & 0.454 & 0.245 & 0.477 & 0.684 & 0.090 & 0.322 & 0.041 & 0.023\\
                \texttt{SDAT} \cite{SDAT} & 0.300$\pm$0.019 & 0.503 & 0.256 & 0.549 & 0.712 & 0.057 & 0.264 & 0.045 & 0.015\\
                \texttt{DANN} \cite{DANN} &  0.306$\pm$0.011 & 0.489 & 0.286 & 0.497 & 0.768 & 0.075 & 0.246 & 0.065 & 0.018\\
                \texttt{MIC} \cite{MIC} & 0.314$\pm$0.009 & 0.531 & 0.264 & 0.556 & 0.719 & 0.069 & 0.291 & 0.059 & 0.020\\  
                \texttt{ELS} \cite{ELS} & 0.318$\pm$0.007 & 0.584 & 0.235 & 0.544 & 0.735 & 0.066 & 0.300 & 0.064 & 0.017\\
                                
                \midrule
                \multicolumn{2}{@{}l}{\textit{SDA baselines:}}\\
                \texttt{CDAN} \cite{CDAN} & 0.570$\pm$0.012 & 0.691 & 0.456 & 0.891 & 0.890 & 0.788 & 0.556 & 0.250 & 0.039\\
                \texttt{ADDA} \cite{ADDA}  & 0.617$\pm$0.012 & 0.726 & 0.467 & 0.923 & 0.883 & 0.825 & 0.590 & 0.306 & 0.220\\
                \texttt{FixBi} \cite{FixBi} & 0.635$\pm$0.011 & 0.710&0.486&0.912&0.888&\textbf{0.861}&0.594&0.422&0.204 \\
                \texttt{MDD} \cite{MDD} & 0.640$\pm$0.015 & 0.712&0.451&0.909&0.860&0.815&0.585&0.532&0.259\\
                \texttt{DANN} \cite{DANN} &  0.640$\pm$0.013 & 0.717&0.423&0.906&0.881&0.827&0.586&0.526&0.251\\
                \texttt{SDAT} \cite{SDAT} & 0.641$\pm$0.016 & 0.746 & 0.458 & 0.928 & 0.909 & 0.838 & 0.610 & 0.444 & 0.193\\
                \texttt{MIC} \cite{MIC} & 0.642$\pm$0.013 & 0.722 & 0.447 & 0.900 & 0.896 & 0.804 & 0.600 & 0.484 & 0.280\\   
                \texttt{CycleGAN} \cite{ju2021leveraging} & 0.651$\pm$0.010 & 0.679&0.471&0.923&0.890&0.817&0.579&0.518&0.331\\
                \texttt{DDC} \cite{DDC} & 0.652$\pm$0.006 & 0.689&0.433&0.888&0.879&0.825&0.616&0.467&\textbf{0.419}\\ 
                \texttt{ELS} \cite{ELS} & 0.656$\pm$0.007 & 0.741 & 0.469 & \textbf{0.934} & \textbf{0.913} & 0.848 & \textbf{0.628} & 0.457 & 0.262\\

                \midrule
                \multicolumn{2}{@{}l}{\textit{Proposed:}} \\
                \texttt{CdCL} & \textbf{0.675}$\pm$0.009 & \textbf{0.747}&0.475&0.932&0.887&0.845&0.614&\textbf{0.581}&0.318\\
                \emph{w/o} CFP & 0.638$\pm$0.010 & 0.732&0.486&0.923&\textbf{0.896}&0.846&0.612&0.431&0.179\\
                \emph{w/o}  \texttt{SbC} & 0.645$\pm$0.026&0.720&0.503&0.920&0.887&0.835&0.615&0.409&0.273 \\
                \emph{w/o}  \texttt{GAP} & 0.662$\pm$0.007 & 0.723&\textbf{0.511}&0.906&0.883&0.832&0.608&0.498&0.331\\
        \bottomrule
            \end{tabular}
        }           
    }
    \vspace{-0.3cm}
\end{table}

\begin{table}[htb]

    \caption{\textbf{Evaluation on WF-MD}.}
    \label{tab:wf-md}
    \renewcommand{\arraystretch}{1.1}
    \centerline{
        \scalebox{0.65}{
                 \begin{tabular}{@{}lrrrrrrrrr@{}}
        \toprule
                \textbf{Method} & \textbf{mAP} & \textbf{DR} &  \textbf{RVO} & \textbf{LP} & \textbf{DM} & \textbf{RD} & \textbf{TF} & \textbf{AMD} & \textbf{MEM}\\
                \midrule  
                
                \multicolumn{2}{@{}l}{\textit{Supervised baselines:}}\\
                \texttt{CFP}                & 0.473$\pm$0.025   & 0.540 &0.659&0.753&0.459&0.509&0.280&0.350&0.234\\
                \texttt{WF}                 & 0.669$\pm$0.003   & 0.760 &0.804&0.926&0.602&0.716&0.619&0.617&0.311\\
                \texttt{WF+}                & 0.680$\pm$0.018   & 0.798 &0.801&0.926&0.657&0.711&0.618&0.587&0.341\\

                \midrule
                \multicolumn{2}{@{}l}{\textit{SDA baselines:}}\\
                \texttt{CDAN} \cite{CDAN}   & 0.613$\pm$0.008   & 0.664 & 0.750 & 0.895 & 0.604 & 0.648 & 0.585 & 0.559 & 0.202\\
                \texttt{MIC} \cite{MIC}     & 0.640$\pm$0.012   & 0.692 & 0.745 & 0.923 & 0.611 & 0.714 & 0.634 & 0.560 & 0.244\\   
                \texttt{FixBi} \cite{FixBi} & 0.652$\pm$0.015   & 0.772 &0.784&0.902&0.630&0.677&0.651&0.595&0.206\\
                \texttt{DDC} \cite{DDC}     & 0.675$\pm$0.002   & 0.756 &0.803&0.920&0.635&0.722&0.630&0.583&0.350\\ 
                \texttt{MDD} \cite{MDD}     & 0.678$\pm$0.007   & 0.743 &0.822&0.910&\textbf{0.658}&0.710&0.631&0.611&0.342\\
                \texttt{ADDA} \cite{ADDA}   & 0.679$\pm$0.007   & 0.786 &0.825&0.937&0.657&0.695&0.617&0.581&0.333\\
                \texttt{SDAT} \cite{SDAT}   & 0.685$\pm$0.011   & 0.764 &0.814 & 0.941 & 0.657 & 0.757 & 0.636 & 0.623 & 0.287 \\
                \texttt{DANN} \cite{DANN}   & 0.686$\pm$0.006   & 0.767 &0.842&0.931&0.594&0.730&0.642&0.604&0.379\\
                \texttt{ELS} \cite{ELS}     & 0.687$\pm$0.006   & 0.772 & 0.824 & \textbf{0.948} & 0.655 & 0.759 & 0.647 & \textbf{0.630} & 0.260\\
                \texttt{CycleGAN} \cite{ju2021leveraging} & 0.688$\pm$0.005&0.773&\textbf{0.850}&0.937&0.644&0.692&0.640&0.581&\textbf{0.384}\\
        
                \midrule
                
                \multicolumn{2}{@{}l}{\textit{Proposed:}} \\
                \texttt{CdCL} & \textbf{0.691}$\pm$0.006 & \textbf{0.813}&0.841&0.921&0.607&\textbf{0.771}&\textbf{0.672}&0.591&0.320\\
                \emph{w/o} CFP & 0.661$\pm$0.009 &0.761&0.773&0.928&0.564&0.728&0.634&0.618&0.285\\
               
                \emph{w/o} \texttt{SbC} & 0.680$\pm$0.007&0.802&0.804&0.905&0.651&0.692&0.642&0.591&0.353 \\

                 \emph{w/o} \texttt{GAP} & 0.684$\pm$0.012 &0.806&0.809&0.918&0.621&0.720&0.662&0.596&0.341\\
        \bottomrule
            \end{tabular}
        }
    }
    \vspace{-0.3cm}
\end{table}


RFMiD has 1,920 color fundus photos for training. The photos, taken by TOPCON 3D OCT-2000 / TRC-NW300 and Kowa VX-10$\alpha$, are annotated \wrt 46 different fundus conditions \cite{RFMiD}. The Tsukazaki Optos Public dataset (TOP) has 13k UWF images taken by Optos 200Tx, annotated with 8 fundus diseases. The two datasets have 3 classes in common, \ie DR, RVO and AMD. In order to match with our setting that labeled UWF images are typically much less than their CFP counterparts, we use one tenth of the TOP training images.

\textbf{Performance metric}. 
We report the commonly used Average Precision (AP) \cite{wei2019laser,mm21-mmmil}. AP per class is calculated, with their mean (mAP) to measure the overall performance.


\textbf{Implementation}. For a fair comparison, we use the following implementation for all methods evaluated in this study, whenever applicable. The backbone network is ImageNet-pretrained EfficientNet-B3 \cite{efficientnet} with pruning \cite{pruned}. 
Subject to our computation capacity (4 Tesla P40 GPUs), 
CFP, WF and UWF images are downsized to $512 \times 512$, $512 \times 512$ and $512 \times 650$, respectively. As the original aspect ratio of all input images were kept unchanged, the input size of UWF images is non-square.
A mini batch has 4 WF/UWF images and their mixup counterparts.
The network optimizer is SGD with cosine annealing strategy \cite{coslr}, an initial learning rate of 1e-3, momentum of 0.95 and weight decay of 1e-4. Early stop occurs if 
the metric does not increase in 10 successive validations. Per method we repeat experiments 5 times, with its mean performance and standard deviation reported.
Ratio $r$ is set to 3 for WF images and 4 for UWF images which will be further discussed in Sec. \ref{abl}.
Our deep learning environment is PyTorch 1.13.

\begin{table}[htb]
    \caption{\textbf{Evaluation on TOP}.}
    \label{tab:top}
    \renewcommand{\arraystretch}{1.1}
    \centerline{
        \scalebox{0.8}{
            \begin{tabular}{@{}lrrrr@{}}
        \toprule
                \textbf{Method} & \textbf{mAP} & \textbf{DR} & \textbf{RVO} & \textbf{AMD} \\
                \midrule   

                \multicolumn{2}{@{}l}{\textit{Supervised baselines:}}\\
                \texttt{CFP} & 0.245$\pm$0.009 &0.571&0.120&0.043\\
                \texttt{UWF+} & 0.640$\pm$0.018 &0.843&0.675&0.403\\
                \texttt{UWF} & 0.647$\pm$0.036 &0.843&0.648&0.451\\
                
                \midrule
                \multicolumn{2}{@{}l}{\textit{SDA baselines:}}\\
                \texttt{DDC} \cite{DDC} & 0.624$\pm$0.015 &0.829&0.643&0.401\\ 
                \texttt{CDAN} \cite{CDAN} & 0.625$\pm$0.046 & 0.828 & 0.640 & 0.407\\
                \texttt{DANN} \cite{DANN} &  0.628$\pm$0.006 &0.830&0.634&0.420\\
                \texttt{MIC} \cite{MIC} & 0.639$\pm$0.030 & 0.817 & 0.664 & 0.437\\   
                \texttt{CycleGAN} \cite{ju2021leveraging} & 0.640$\pm$0.018&0.827&0.639&0.453\\
                \texttt{MDD} \cite{MDD} & 0.644$\pm$0.012 &0.839&0.640&0.453\\
                \texttt{FixBi} \cite{FixBi} & 0.653$\pm$0.024 &0.845&0.670&0.445\\
                \texttt{ELS} \cite{ELS} & 0.660$\pm$0.031 & 0.845 & 0.661 & 0.472\\
                \texttt{SDAT} \cite{SDAT} & 0.661$\pm$0.021 & 0.846 & \textbf{0.679} & 0.458\\
                \texttt{ADDA} \cite{ADDA}  & 0.663$\pm$0.022 &0.845&0.671&0.472\\
                                
                \midrule
                \multicolumn{2}{@{}l}{\textit{Proposed:}}\\
                \texttt{CdCL} & \textbf{0.678}$\pm$0.032 &\textbf{0.857}&0.658&\textbf{0.519}\\
                \emph{w/o} CFP & 0.652$\pm$0.024 &0.847&0.662&0.446\\
                
                \emph{w/o} \texttt{SbC} & 0.662$\pm$0.020&0.852&\textbf{0.679}&0.455 \\

                \emph{w/o} \texttt{GAP} & 0.677$\pm$0.032 & 0.855&0.666&0.506\\
                
        \bottomrule
            \end{tabular}
        }           
    }
    \vspace{-0.3cm}
\end{table}

\textbf{Baselines}. We have three relatively straightforward supervised baselines. That is, \texttt{CFP} trained exclusively on the CFP data, \texttt{WF} / \texttt{UWF} trained on the relatively limited  WF / UWF samples, and \texttt{WF+} / \texttt{UWF+} which uses \texttt{CFP} for weight initialization. 
Moreover, we compare with \texttt{CycleGAN} \cite{ju2021leveraging} designed specifically for UWF image classification. We also include the following generic methods for comparison, \ie \texttt{DDC} \cite{DDC}, \texttt{DANN} \cite{DANN}, \texttt{ADDA} \cite{ADDA}, \texttt{CDAN} \cite{CDAN}, \texttt{MDD} \cite{MDD}, \texttt{FixBi} \cite{FixBi}, \texttt{SDAT} \cite{SDAT}, \texttt{ELS} \cite{ELS} and \texttt{MIC} \cite{MIC}. Note that these methods were typically evaluated in the context of unsupervised domain adaptation, where the target domain is assumed to be unlabeled. So for a fair comparison, we further provide SDA versions of these method, in which a supervised loss concerning WF/UWF training images is added, improving their performance to a large extent.

\subsection{Results and Analysis} \label{ssec:results}

\subsubsection{Overall comparison}
AP scores of different methods on UWF-MD, WF-MD and TOP are given in Table \ref{tab:uwf-md}, \ref{tab:wf-md} and \ref{tab:top}, respectively. The best baseline per dataset varies as follows: ELS on UWF-MD with mAP of 0.656, CycleGAN on WF-MD with mAP of 0.688, and ADDA on TOP with mAP of 0.663. The proposed \texttt{CdCL} method consistently surpasses the best baselines, with mAP of 0.675 on UWF-MD, 0.691 on WF-MD and 0.678 on TOP. 

As shown in Table \ref{tab:uwf-md}, there is a notable disparity between UDA baselines and SDA baselines. Hence, subsequent experiments do not include the UDA baselines.

\begin{figure}[tb!]
    \centerline{
        \subcaptionbox{\label{fig:plot1}}{
        \includegraphics[height=0.25\columnwidth]{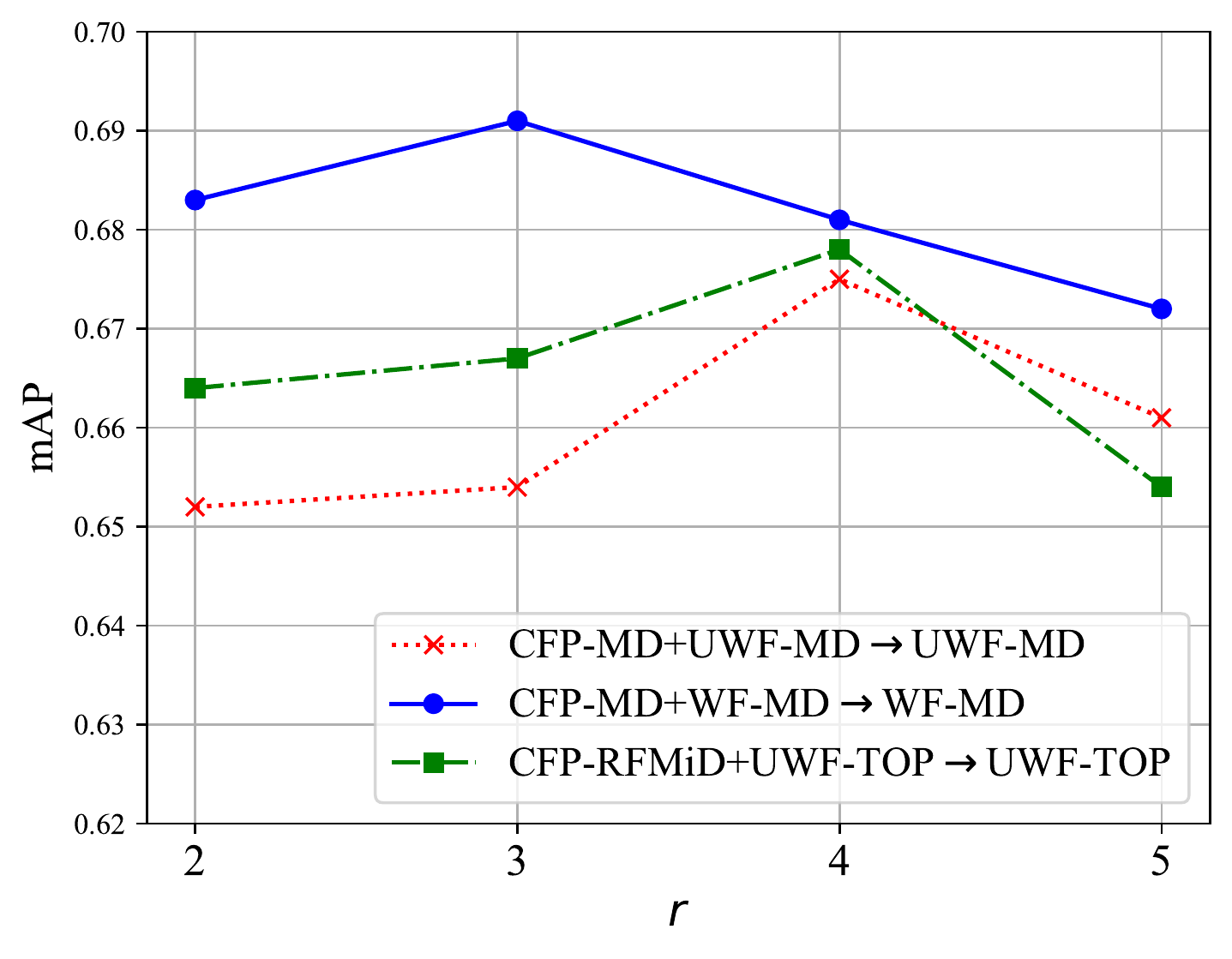}}
            
        \subcaptionbox{\label{fig:plot2}}{
        \includegraphics[height=0.25\columnwidth]{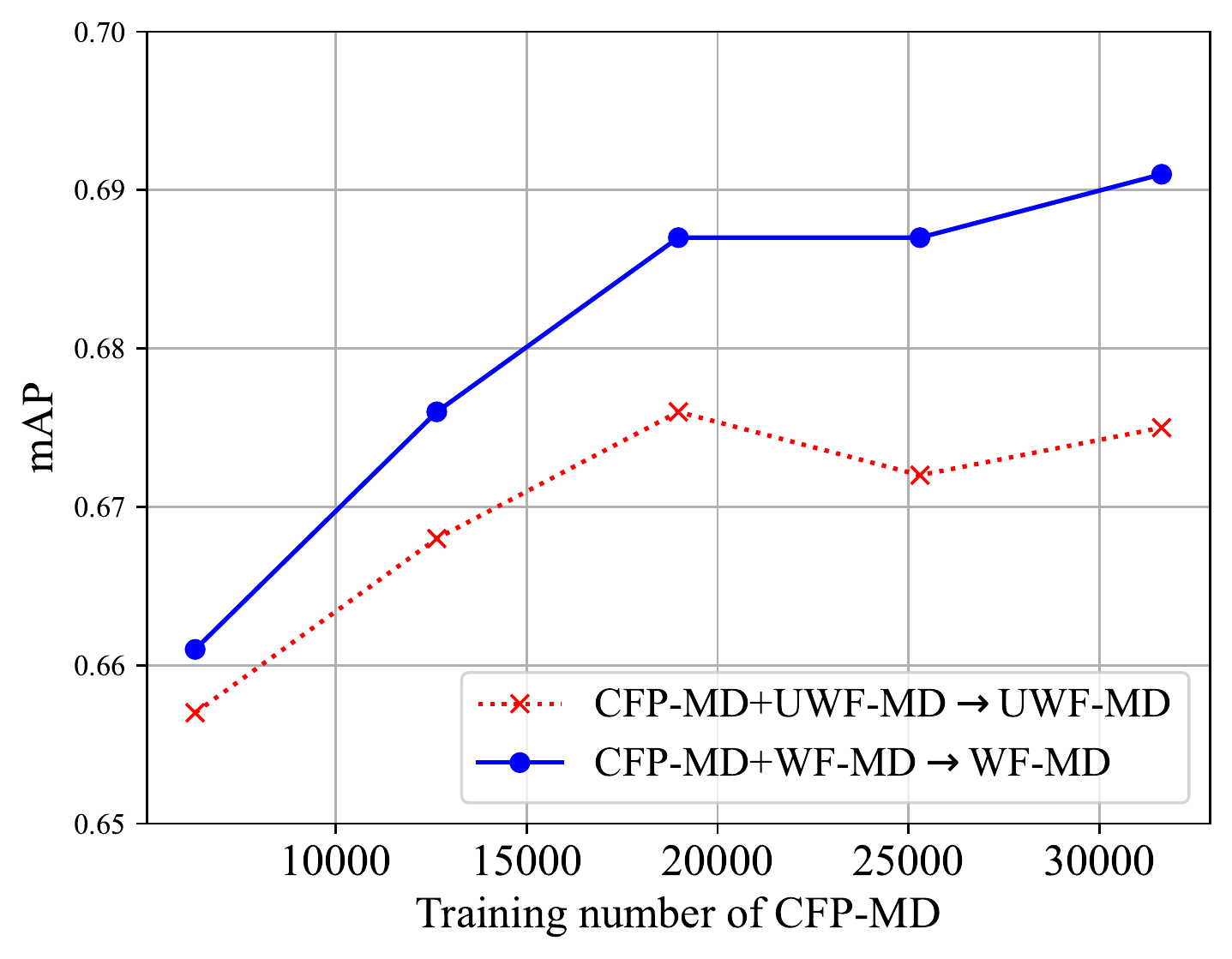}}
            
        \subcaptionbox{\label{fig:plot3}}{
        \includegraphics[height=0.25\columnwidth]{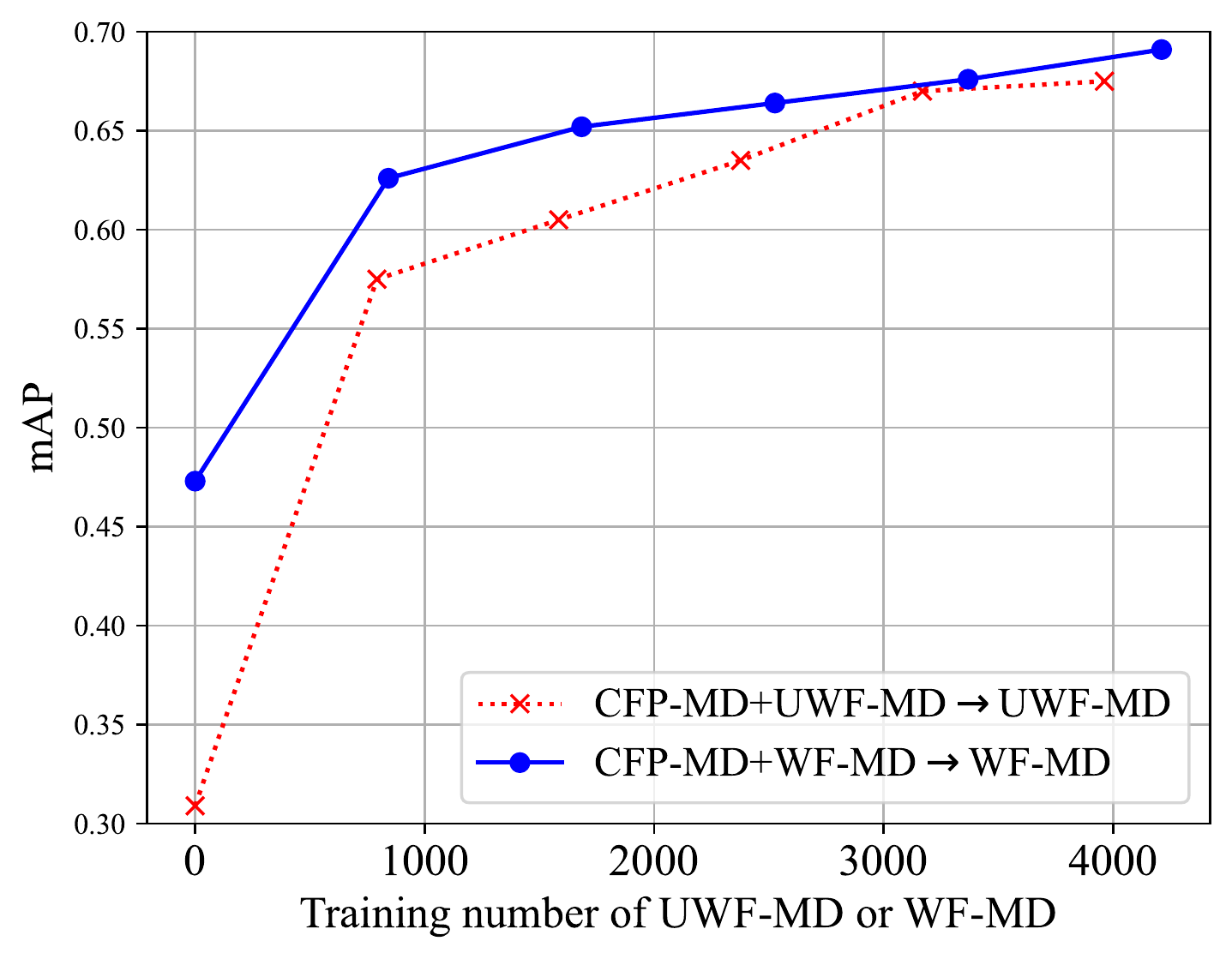}}}
            
            
    \caption{\textbf{The performance of \texttt{CdCL} in varied training settings: (a) $r$, (b) the amount of CFP images and (c) the amount of UWF/WF images.}}
    \label{fig:plot}
    \vspace{-0.3cm}
\end{figure}


Notice that in contrast to UWF-MD and TOP, our improvement on WF-MD seems marginal. We attribute this to the fact that compared to UWF images, WF images captured by the ZEISS Clarus500 camera are more close to CFP images in terms of their visual appearance and FoV, see Fig. \ref{fig:example}, meaning relatively smaller domain divergence. This is also confirmed by the result that the performance gap between \texttt{CFP} and \texttt{WF} is the smallest, \ie 0.669-0.473=0.226, while the corresponding figures between \texttt{CFP} and \texttt{UWF} are 0.622-0.309=0.313 and 0.647-0.245=0.402.

    

    

        

\subsubsection{Ablation study}
\label{abl}
\textbf{Effects of different components}. Recall that we add two MHSA blocks for scale-bias correction (SbC). With the extra modules added, the learning capacity of the network naturally grows. To which shall we attribute the performance gain, the stronger network or the cross-domain use of CFPs? To resolve such uncertainty, we re-train the network without using CFP training data. Clear performance drop can be consistently observed: UWF-MD 0.675 $\rightarrow$ 0.638, WF-MD 0.691 $\rightarrow$ 0.661, and TOP 0.678 $\rightarrow$ 0.652. The result allows us to safely conclude that the cross-domain use of CFPs is necessary. 

Furthermore, we try another configuration with the SbC module removed. Noticeable performance loss is also observed on all the three datasets: UWF-MD 0.675 $\rightarrow$ 0.645, WF-MD 0.691 $\rightarrow$ 0.680, and TOP 0.678 $\rightarrow$ 0.662. The importance of SbC for \texttt{CdCL} is thus verified. 

We also try to make prediction exclusively on the scale-bias corrected feature. This run is denoted as \emph{w/o} \texttt{GAP} in the result tables. Its lower performance suggests the feature by GAP remains beneficial.

\textbf{Effects of ratio $r$}.
We set different $r$ to \texttt{CdCL} to investigate its impact on the performance. As shown in Fig.~\ref{fig:plot1}, we found that the optimal value of $r$ is 4 for UWF images and 3 for WF images. 
Given that UWF images have a larger FoV than WF images, the lesions in UWF images are smaller than those in WF images. 
Consequently, a smaller scope of local features is required for UWF images, resulting in a larger value for $r$.

\textbf{Effects of training data}
We study the influence of the amount of CFP and WF/UWF images used for training. 
As shown in Fig.~\ref{fig:plot2}, with the full training data of WF/UWF images used, using about 20k CFP images (60\% of the full set) is adequate for relatively high performance. 

On the other hand, under circumstance that the full training data of CFP images is used, we evaluate the performance of \texttt{CdCL} with partial training data of WF/UWF images. As shown in Fig.~\ref{fig:plot3}, a relatively small amount of WF/UWF images (20\% of the full set) can bring in substantial increasing in performance. In addition, as the amount of WF/UWF images increases, the performance continues to improve.

\section{Conclusions and Remarking} \label{sec:conc}

Given a relatively small amount of labeled samples from a target domain, \ie wide-field (WF) or ultra-wide-field (UWF) imaging, 
this paper studies how to leverage a larger amount of existing labeled color fundus photos (CFPs) for recognizing multiple fundus diseases on the target-domain images.
Extensive experiments on three datasets covering both UWF and WF images support our conclusions as follows.
The proposed Cross-domain Collaborative Learning  method is effective, beating all baseline methods in consideration on all the three datasets. For an effective cross-domain use of CFP data, scale-bias correction (SbC) on the original CNN features is necessary. SbC might also be useful for cross-domain learning in other contexts.  


\balance
\bibliographystyle{IEEEtran}
\bibliography{refs}

\end{document}